**Solid-liquid equilibria and excess enthalpies in binary mixtures of propiophenone with some aliphatic amides**

Patryk Sikorski [b], Ana Cobos Huerga [a], Juan Antonio González [a], Marek Królikowski [b], Tadeusz Hofman[b,*]

[a] Warsaw University of Technology, Faculty of Chemistry, ul. Noakowskiego 3, 00-664 Warszawa, Poland
[b] Universidad de Valladolid, Dpto Física Aplicada, Facultad de Ciencias, Paseo de Belén 7, 47011 Valladolid, Spain

**ABSTRACT**

The solid-liquid equilibria and excess enthalpies of the binary systems of propiophenone and {N-methylformamide, or N,N-dimethylformamide, or N,N-dimethylacetamide, or N-methyl-2-pyrrolidone} were measured. The SLE data were determined by the cloud-point method, partly supplemented by the DSC technique. The excess enthalpies were measured by the titration calorimetry at 298.15 K and 308.15 K for all the systems and additionally at 293.15 K and 318.15 K for the system with N-methylformamide, and at 293.15 K for the system with N,N-dimethylformamide. The activity coefficients of both components as a function of concentration and temperature were calculated by the correlation which included both types of data. The results of the prediction by the modified UNIFAC model were compared with the experimental data. The observed trends and differences between the systems were discussed.

**1. Introduction**

We present a continuation of research on the thermodynamic properties of systems composed of compounds containing amide and carbonyl groups in which strong and various molecular interactions appear. In our previous paper, the binary systems with acetophenone and {N-methylformamide, or N,N-dimethylformamide, or N,N-dimethylacetamide, or N-methyl-2-pyrrolidone} were studied through the solid-liquid equilibria and excess enthalpy measurements [1]. We decided to extend this work to systems with propiophenone as the influence of the aliphatic chain adjacent to the carbonyl group can be investigated.

Propiophenone is used in the industry less frequently than acetophenone, nevertheless its application still is quite wide – it is used as a substrate in pharmaceutical synthesis and as flavoring agent [2]. It is a highly polar compound with a dipole moment estimated to be slightly less than 3 D, which is the value for acetophenone [3]. The second components, i.e. (N-methylformamide - NMF, N,N-dimethylformamide - DMF, N,N-dimethylacetamide - DMA) belong to the group of aliphatic amides while (N-methyl-2-pyrrolidone – NMP) is a lactam. They are very polar compounds with a dipole moment close to 4 D [3]. N-methylformamide is known for an extremely high dielectric constant of 189.0 at 293.2 K [3].

In our previous publication we emphasized an interest on the problem of the group definition in the group-contribution approach [1]. A strongly polar group with some adjacent nonpolar ones usually is treated as one macro-group which requires unique energetic parameters. If they are not available a division into simpler groups may be necessary although the outcome of this approach is unclear. Formally, the amide group could be defined as two separate groups – amine and carbonyl ones. The thermodynamic data reported in the literature concerning the systems with propiophenone and NMF, DMF, DMA or NMP, or similar amides are extremely rare and present only two data sets of densities, excess volumes, viscosity and speed of sound, measured at the same laboratory for the system propiophenone + N-methylacetamide at 308.15 K [4] [5].

## 2. Materials

Propiophenone, N-methylformamide, N,N-dimethylformamide, N,N-dimethylacetamide and N-methyl-2-pyrrolidone were purchased in Sigma-Aldrich company. N-methylformamide  was purified by distillation what gave the mass purity confirmed by the GC equal to 0.996. Additional information is given in Table 1.

Table 1. The compounds used in measurements

| Compound | CAS | Purity (mass)[a] | Water content/ppm |
|---|---|---|---|
| Propiophenone | 93-55-0 | 0.999[a] | 10[b] |
| N-methylformamide | 123-39-7 | 0.996[ac] | 260[b] |
| N,N-dimethylformamide | 68-12-2 | 0.998[d] | < 50[a] |
| N,N-dimethylacetamide | 127-19-5 | 0.998[d] | < 50[a] |
| N-methyl-2-pyrrolidone | 872-50-4 | 0.995[a\d] | < 50[a] |

[a]GC. [b] Measured by the Fischer method. [c] After distillation. [d]Declared by the producer.

## 3. Experimental

### 3.1. Solid-liquid equilibria

The cloud-point technique was used to determine solid-liquid equilibria. Its main idea is to slowly increase temperature of a sample of known composition up to appearance of homogenous mixture which is equivalent to the solubility temperature. Numerous variants of this method differ mainly as to the detection of this temperature. The original idea is believed to be invented by Alekseev in XIX century yet [6]. The details of the procedure are given elsewhere [1] and only the most important issues will be presented here. The schematic drawing of the apparatus is shown in Fig. 1.

The light intensity (luminance) as a function of temperature was a relationship measured directly. The latter parameter was slowly increased, which resulted in a steady increase in luminance, until crystals disappeared, from then the luminance was independent of the temperature. This solubility temperature corresponds to the break in the luminance *versus* temperature curve. In the vicinity of the solubility point, the heating rate did not exceed  2.5 K·h$^{-1}$.

Standard uncertainty of temperature was estimated to be about 0.5 K. The main source of error, apart from the dynamic character of the method, was the determination of the solubility point which was taken as an intersection of two straight lines – one increasing and one parallel to the temperature axis.

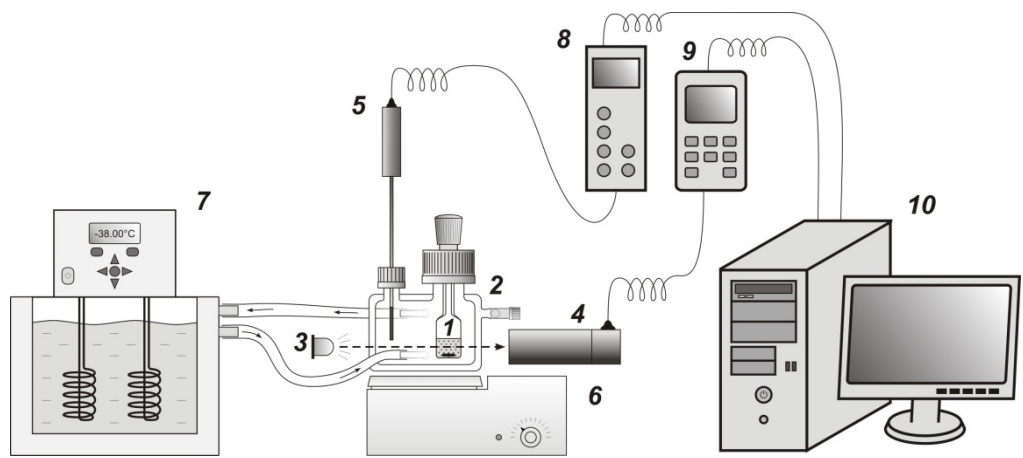

Fig. 1
The cloud-point technique apparatus: (1) equilibrium cell, (2) insulated vessel; (3) source of light; (4) probe measuring the intensity of light; (5) temperature probe; (6) magnetic stirrer; (7) cryostat/thermostat; (8) temperature measuring unit; (9) detector of light; (10) CPU.

This method has a technical limitation due to cryostat efficiency which did not allow temperature to drop below -30 °C. For some systems and some ranges of composition, if such low solubility temperatures were encountered, the solubility data were measured by the Differential Scanning Calorimetry method. For this purpose the Mettler Toledo DSC1 STAR® apparatus was used which was calibrated with the scan rate of 5 K·min$^{-1}$ with water, $n$-heptane, $n$-octane and indium sample of 0.999999 mole fraction purity. The measured samples were heated with the rate of 1 K·min$^{-1}$. The eutectic (onset temperature) and solubility (peak minimum) temperatures were calculated by the STAR software. The standard combined uncertainty of temperature was estimated through data scattering to be at least 1 K.

The standard uncertainty of mole fractions for both methods was about 1·10$^{-4}$. The details of the experimental procedure are given in our previous paper [1].

The melting temperatures of pure compounds and the corresponding enthalpies of fusion are given in Table 2. The enthalpy of melting of propiophenone was measured in this work. Its value is slightly higher than that of acetophenone [7].

The determined solid-liquid equilibria data are collected in Table 3 and illustrated in Figures 2-5. The solubility temperatures measured by the DSC technique are designated by a superscript "c". The three solubility temperatures measured for the propiophenone + DMF system at the lowest concentration of propiophenone could not be distinguished from the eutectic temperature. We interpreted them as corresponding to the eutectic temperature due to their relative constancy at 210 K.

Table 2. Melting temperatures and enthalpies of fusion of pure compounds.

| | $T_m$/K | $\Delta H_m$/ kJ·mol$^{-1}$ | reference |
|---|---|---|---|
| Propiophenone | 291.8 | | [8] |
| | 291.4[a] | 19.3 | this work[c] |
| N-methylformamide | 270.6 | 10.44 | [9] |
| | 270.8 | | this work[b] |
| N,N-dimethylformamide | 212.9 | 8.95 | [10] |
| | 211.7[a] | | this work[c] |
| N,N-dimethylacetamide | 253.2 | 8.2 | [11] |
| | 254.2 | 10.2 | [10] |
| | 251.4 | 10.4 | [9] |
| | 250.5 | | this work[b] |
| N-methyl-2-pyrrolidone | 248.5 | 18.1 | [11] |
| | 248.3 | | this work[b] |

[a] Melting temperature measured by the DSC method. [b] Standard uncertainty $u(T)$ = 0.5 K. [c] Standard uncertainty u($T$) = 1 K.

Table 3. Solid-liquid equilibria of the propiophenone (1) + {NMF, or DMF, or DMA, or NMP} (2) systems.[a]

| $x_1$ | $T$/K | solid phase | $\gamma_1$ | | $\gamma_2$ | |
|---|---|---|---|---|---|---|
| | | | exp[b] | calc | exp[b] | calc |
| Propiophenone (1) + N-methylformamide (2) | | | | | | |
| 0 | 270.8 | 2 | | | 1 | 1 |
| 0.0463 | 268.0 | 2 | | 4.857 | 0.999 | 1.006 |
| 0.0956 | 266.1 | 2 | | 3.882 | 1.019 | 1.024 |
| 0.1233 | 265.3 | 1 | 3.704 | 3.658 | | 1.031 |
| 0.1494 | 267.2 | 1 | 3.253 | 3.324 | | 1.045 |
| 0.2007 | 272.1 | 1 | 2.832 | 2.822 | | 1.080 |
| 0.2514 | 275.0 | 1 | 2.474 | 2.464 | | 1.121 |
| 0.2976 | 277.1 | 1 | 2.228 | 2.216 | | 1.165 |
| 0.3494 | 279.0 | 1 | 2.009 | 1.999 | | 1.222 |
| 0.4501 | 281.7 | 1 | 1.689 | 1.694 | | 1.359 |
| 0.4011 | 280.6 | 1 | 1.835 | 1.828 | | 1.287 |
| 0.4989 | 282.8 | 1 | 1.573 | 1.580 | | 1.444 |
| 0.5548 | 283.9 | 1 | 1.460 | 1.468 | | 1.565 |
| 0.6031 | 284.4 | 1 | 1.363 | 1.382 | | 1.697 |
| 0.6527 | 285.5 | 1 | 1.300 | 1.303 | | 1.871 |
| 0.7039 | 286.0 | 1 | 1.222 | 1.229 | | 2.110 |
| 0.7529 | 287.0 | 1 | 1.176 | 1.167 | | 2.422 |
| 0.8025 | 287.7 | 1 | 1.125 | 1.112 | | 2.861 |
| 0.8510 | 288.7 | 1 | 1.091 | 1.067 | | 3.476 |
| 0.8983 | 289.7 | 1 | 1.062 | 1.033 | | 4.355 |
| 0.9499 | 290.7 | 1 | 1.033 | 1.009 | | 5.835 |
| 1 | 291.4 | 1 | 1 | 1 | | |

| Propiophenone (1) + N,N-dimethylformamide (2) | | | | | |
|---|---|---|---|---|---|
| 0 | 211.7[c] | 2 | | | 1 |
| 0.0498 | 209.9[c] | 1+2 | | | |
| 0.1003 | 210.1[c] | 1+2 | | | |
| 0.1749 | 209.8[c] | 1+2 | | | |
| 0.1749 | 242.2c | 1 | 1.134 | 1.194 | 1.013 |
| 0.2023 | 248.4 | 1 | 1.245 | 1.185 | 1.016 |
| 0.2488 | 253.3 | 1 | 1.213 | 1.169 | 1.021 |
| 0.2982 | 256.7 | 1 | 1.142 | 1.149 | 1.028 |
| 0.3427 | 260.4 | 1 | 1.130 | 1.132 | 1.035 |
| 0.4020 | 264.2 | 1 | 1.096 | 1.110 | 1.048 |
| 0.4441 | 267.2 | 1 | 1.094 | 1.095 | 1.058 |
| 0.4983 | 270.2 | 1 | 1.074 | 1.078 | 1.073 |
| 0.5495 | 272.7 | 1 | 1.054 | 1.063 | 1.089 |
| 0.6009 | 275.3 | 1 | 1.044 | 1.049 | 1.107 |
| 0.6500 | 277.2 | 1 | 1.023 | 1.038 | 1.127 |
| 0.7030 | 279.5 | 1 | 1.013 | 1.027 | 1.151 |
| 0.7467 | 281.4 | 1 | 1.009 | 1.020 | 1.173 |
| 0.7991 | 283.5 | 1 | 1.002 | 1.012 | 1.201 |
| 0.8577 | 286.7 | 1 | 1.023 | 1.006 | 1.236 |
| 0.8986 | 288.6 | 1 | 1.030 | 1.003 | 1.263 |
| 0.9449 | 290.0 | 1 | 1.018 | 1.001 | 1.295 |
| 1 | 291.4 | 1 | 1 | 1 | |

| Propiophenone (1) + N,N-dimethylacetamide (2) | | | | | | |
|---|---|---|---|---|---|---|
| 0 | 250.5 | 2 | | | 1 | 1 |
| 0.0546 | 246.3 | 2 | | 0.206 | 0.973 | 0.961 |
| 0.0998 | 240.4[c] | 2 | | 0.443 | 0.904 | 0.903 |
| 0.1498 | 233.9[c] | 2 | | 0.720 | 0.831 | 0.843 |
| 0.2005 | 240.1[c] | 1 | 0.909 | 0.890 | | 0.807 |
| 0.2435 | 246.8 | 1 | 0.973 | 0.968 | | 0.789 |
| 0.3001 | 252.0 | 1 | 0.959 | 0.994 | | 0.782 |
| 0.3508 | 257.1 | 1 | 0.985 | 0.987 | | 0.786 |
| 0.3942 | 259.7 | 1 | 0.959 | 0.978 | | 0.791 |
| 0.4434 | 264.4 | 1 | 1.000 | 0.975 | | 0.794 |
| 0.4990 | 268.0 | 1 | 1.000 | 0.984 | | 0.789 |
| 0.5541 | 272.0 | 1 | 1.023 | 1.000 | | 0.775 |
| 0.6006 | 274.0 | 1 | 1.004 | 1.016 | | 0.761 |
| 0.6465 | 276.8 | 1 | 1.016 | 1.027 | | 0.747 |
| 0.7045 | 279.8 | 1 | 1.020 | 1.033 | | 0.740 |
| 0.7504 | 281.8 | 1 | 1.016 | 1.030 | | 0.747 |
| 0.7920 | 283.8 | 1 | 1.020 | 1.023 | | 0.767 |
| 0.8532 | 286.6 | 1 | 1.026 | 1.009 | | 0.817 |
| 0.8978 | 288.5 | 1 | 1.028 | 1.002 | | 0.861 |
| 0.9485 | 290.6 | 1 | 1.031 | 0.999 | | 0.890 |
| 1 | 291.4 | 1 | 1 | 1 | | |

Propiophenone (1) + N-methyl-2-pyrrolidone (2)

| $x_1$ | $T$/K | | | | | |
|---|---|---|---|---|---|---|
| 0 | 248.3 | 2 | | | 1 | 1 |
| 0.0520 | 248.1 | 2 | | 1.094 | 1.047 | 1.004 |
| 0.0980 | 246.2 | 2 | | 0.995 | 1.029 | 1.011 |
| 0.1546[c] | 242.5[c] | 2 | | 0.913 | 0.959 | 1.023 |
| 0.1546[c] | 236.4[c] | 1+2 | | | | |
| 0.2498[c] | 242.9[c] | 1 | 0.816 | 0.812 | | 1.059 |
| 0.3041 | 247.5 | 1 | 0.801 | 0.822 | | 1.061 |
| 0.3506 | 252.2 | 1 | 0.827 | 0.842 | | 1.055 |
| 0.3962 | 257.4 | 1 | 0.881 | 0.867 | | 1.043 |
| 0.4577 | 263.2 | 1 | 0.931 | 0.907 | | 1.017 |
| 0.5008 | 266.7 | 1 | 0.955 | 0.935 | | 0.994 |
| 0.5455 | 269.7 | 1 | 0.966 | 0.962 | | 0.969 |
| 0.5995 | 273.5 | 1 | 0.990 | 0.991 | | 0.938 |
| 0.6501 | 276.3 | 1 | 0.995 | 1.011 | | 0.913 |
| 0.6954 | 278.9 | 1 | 1.006 | 1.022 | | 0.896 |
| 0.7481 | 281.7 | 1 | 1.016 | 1.028 | | 0.888 |
| 0.7925 | 283.9 | 1 | 1.022 | 1.027 | | 0.895 |
| 0.8452 | 286.3 | 1 | 1.027 | 1.021 | | 0.925 |
| 0.8948 | 288.1 | 1 | 1.020 | 1.013 | | 0.985 |
| 0.9463 | 290.2 | 1 | 1.023 | 1.004 | | 1.093 |
| 1 | 291.4 | 1 | 1 | 1 | | |

[a] Standard uncertainties $u$ are: $u(x_1) = 1 \cdot 10^{-4}$, $u(T) = 0.5$ K for the cloud-point technique and 1 K for the DSC. [b] Experimental activity coefficients calculated by means of eqns. (4,5). [c] Data obtained by the DSC method.

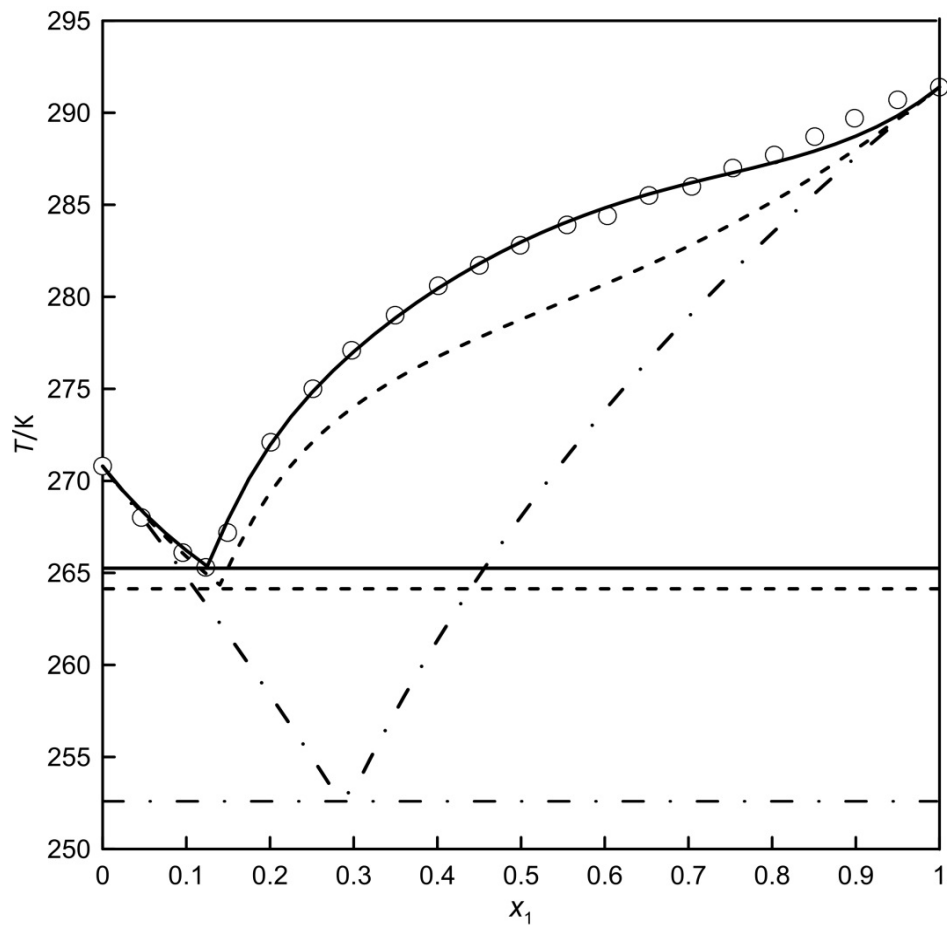

Fig. 2

Solid-liquid equilibria for the propiophenone (1) + N-methylformamide (2) system: (○) – experimental, solid line – calculated by eqn. (5) with the parameters given in Table 6, dashed line – predicted by the modified UNIFAC model, dashed dotted line – ideal solubility.

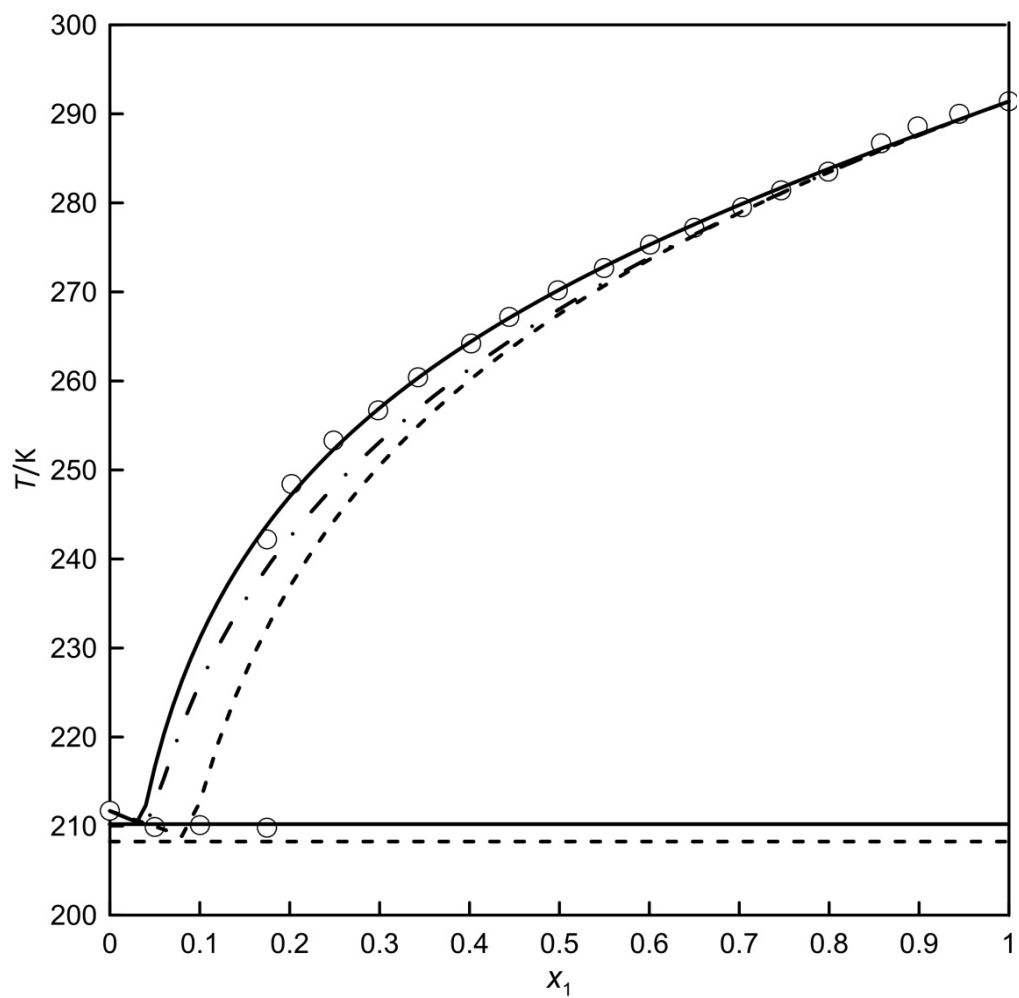

Fig. 3
Solid-liquid equilibria for the propiophenone (1) + N,N-dimethylformamide (2) system: (○) – experimental, solid line – calculated by eqn. (5) with the parameters given in Table 6, dashed line – predicted by the modified UNIFAC model, dashed dotted line – ideal solubility.

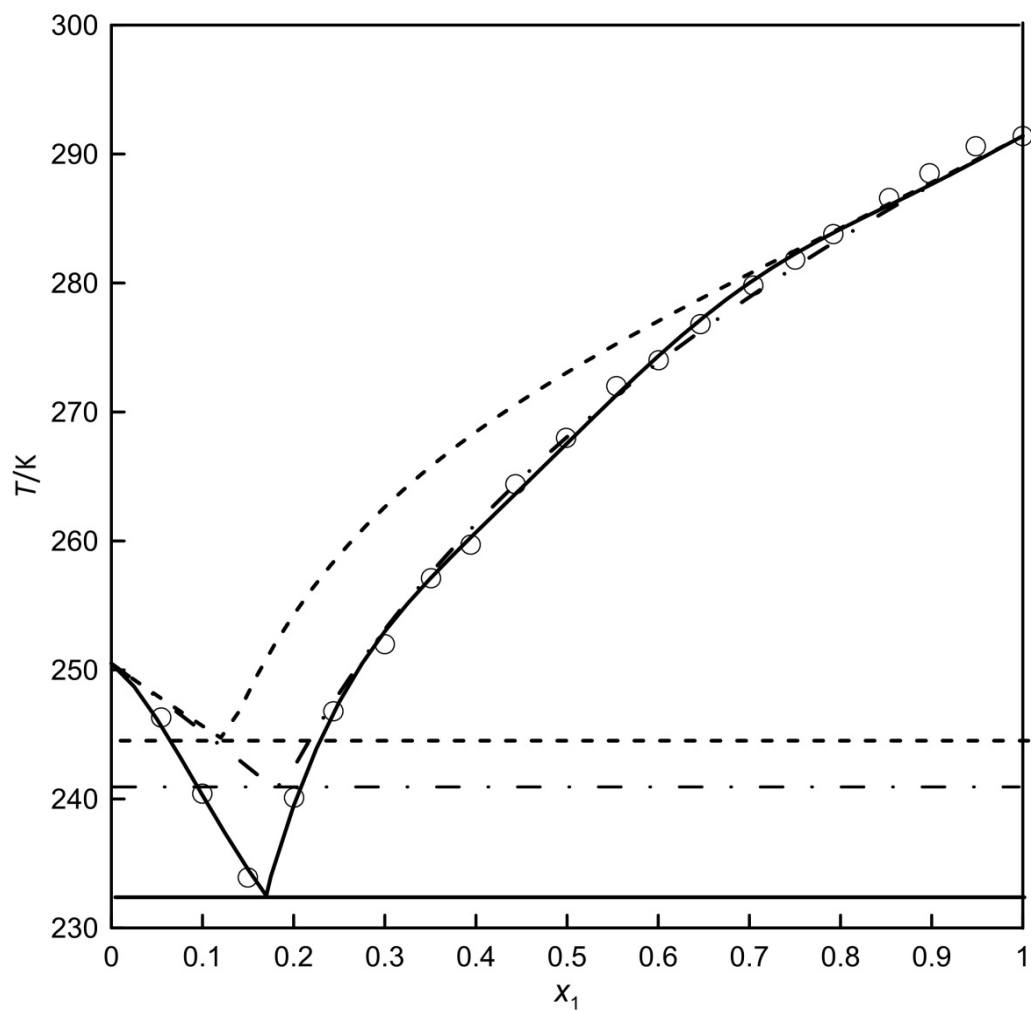

Fig. 4

Solid-liquid equilibria for the propiophenone (1) + N,N-dimethylacetamide (2) system: (○) – experimental, solid line – calculated by eqn. (5) with the parameters given in Table 6, dashed line – predicted by the modified UNIFAC model, dashed dotted line – ideal solubility.

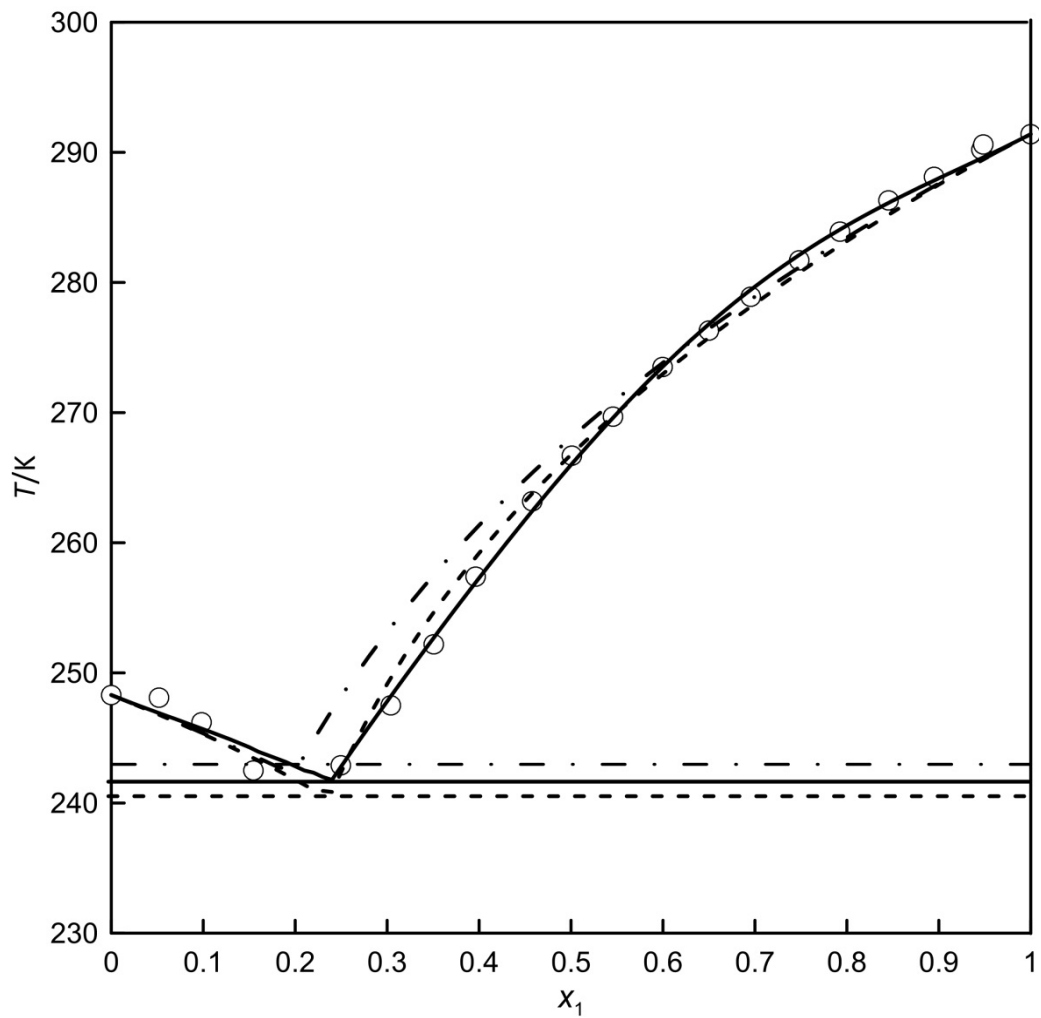

Fig. 5

Solid-liquid equilibria for the propiophenone (1) + N-methyl-2-pyrrolidone system (2): (○) – experimental, solid line – calculated by eqn. (5) with the parameters given in Table 6, dashed line – predicted by the modified UNIFAC model, dashed dotted line – ideal solubility.

### 3.2. Excess Enthalpy

The excess enthalpies of mixing were measured by the TAM III calorimeter using the isothermal titration calorimetry. The details of experimental procedure was described in our previous paper [1]. Although the temperature in the bath was kept constant within ±0.1 mK, the standard uncertainty of this parameter we estimated to be about 0.03 K due to non-compensated heat effects resulting from mixing. The amounts of pure compounds mixed together to form a mixture were determined volumetrically and recalculated into mole fractions using known densities. The total data set consisted of two series – for each one successive amounts of one component were added to the fixed amount of the second component placed in the titration cell. For the systems of propiophenone + {DMA, or NMP) the measurements were carried out at 298.15 K and 308.15 K, for the propiophenone + NMF at (293.15, 298.15, 308.15, and 318.15) K and for the propiophenone + DMF at (293.15, 298.15, and 308.15) K.

The excess enthalpy was calculated as a superposition of heat effects resulting from successive injections of a certain volume of one component into a titration cell, according to the following equation

$$H_i^E = \frac{\sum_{j=1}^{i} \delta q_j}{n_1 + \sum_{j=1}^{i} \Delta n_{2j}} \qquad (1)$$

Where $\delta q_j$ is the heat effect during the jth injection, $n_1$ is the number of moles of the compound 1 in the titration cell before the titration starts and $\Delta n_{2j}$ is the number of moles of the compound 2 injected into the titration cell during the $j$th tritration. The relative standard uncertainty of excess enthalpy is estimated to be about 0.5%.

The measured enthalpies are collected in Table 4. Additionally, two dependences (for the propiophenone + NMF and propiophenone + DMA) are illustrated in Fig. 6 and 7. The former reveals a high endothermic effect while the latter one – medium exothermic. All the data were measured for the first time and cannot be compared to the literature data. They seem to be consistent with our data measured for analogous systems in which propiophenone was replaced by acetophenone [1].

Table 4. Excess enthalpies of the propiophenone + {N-methylformamide or N,N-dimethylformamide, or N,N-dimethylacetamide, or N-methyl-2-pyrrolidone} mixtures.[a]

| $x_1$ | $H_m^E$/ J·mol$^{-1}$ | $x_1$ | $H_m^E$/ J·mol$^{-1}$ | $x_1$ | $H_m^E$/ J·mol$^{-1}$ |
|---|---|---|---|---|---|
| \multicolumn{6} Propiophenone (1) + N-methylformamide (2), $T$ = 293.15 K | | | | | |
| 0.0300 | 61.9 | 0.3404 | 472.2 | 0.6708 | 562.9 |
| 0.0600 | 117.9 | 0.3704 | 496.4 | 0.7008 | 549.7 |
| 0.0899 | 171.0 | 0.4005 | 517.1 | 0.7309 | 531.7 |
| 0.1199 | 218.6 | 0.4305 | 534.8 | 0.7609 | 509.1 |
| 0.1499 | 266.9 | 0.4606 | 549.5 | 0.7909 | 480.9 |
| 0.1799 | 309.5 | 0.4906 | 561.3 | 0.8207 | 446.3 |
| 0.2099 | 348.8 | 0.5207 | 569.8 | 0.8506 | 403.9 |
| 0.2399 | 383.3 | 0.5507 | 575.0 | 0.8806 | 352.7 |
| 0.2698 | 417.7 | 0.5807 | 577.9 | 0.9104 | 289.7 |
| 0.2998 | 447.5 | 0.6107 | 576.7 | 0.9404 | 212.8 |
| 0.3048 | 452.3 | 0.6407 | 571.6 | 0.9702 | 115.0 |
| \multicolumn{6} Propiophenone (1) + N-methylformamide (2), $T$ = 298.15 K | | | | | |
| 0.0200 | 43.3 | 0.2398 | 396.5 | 0.5605 | 595.9 |
| 0.0400 | 84.4 | 0.2598 | 419.2 | 0.6005 | 595.7 |
| 0.0600 | 123.4 | 0.2798 | 440.6 | 0.6406 | 589.2 |
| 0.0799 | 160.4 | 0.2998 | 460.7 | 0.6805 | 575.9 |
| 0.0999 | 195.6 | 0.3102 | 470.0 | 0.7205 | 554.7 |
| 0.1198 | 229.0 | 0.3202 | 479.1 | 0.7605 | 524.8 |
| 0.1398 | 260.7 | 0.3603 | 511.7 | 0.8005 | 484.7 |
| 0.1598 | 290.9 | 0.4004 | 539.1 | 0.8405 | 432.6 |
| 0.1798 | 319.5 | 0.4404 | 561.6 | 0.8803 | 364.9 |
| 0.1998 | 346.6 | 0.4804 | 578.6 | 0.9202 | 276.9 |
| 0.2198 | 372.3 | 0.5205 | 590.0 | 0.9602 | 160.9 |
| \multicolumn{6} Propiophenone (1) + N-methylformamide (2), $T$ = 308.15 K | | | | | |
| 0.0200 | 45.4 | 0.2398 | 424.0 | 0.6005 | 613.1 |
| 0.0400 | 88.1 | 0.2598 | 447.5 | 0.6406 | 605.1 |
| 0.0600 | 130.9 | 0.2798 | 470.2 | 0.6806 | 590.0 |
| 0.0799 | 170.0 | 0.2998 | 491.8 | 0.7206 | 567.2 |
| 0.0999 | 208.2 | 0.3203 | 501.9 | 0.7605 | 535.2 |
| 0.1199 | 245.1 | 0.3603 | 534.1 | 0.8005 | 492.8 |
| 0.1399 | 279.0 | 0.4003 | 561.2 | 0.8405 | 438.5 |
| 0.1599 | 311.9 | 0.4404 | 582.9 | 0.8804 | 369.3 |
| 0.1799 | 341.8 | 0.4804 | 599.0 | 0.9203 | 279.7 |
| 0.1999 | 371.3 | 0.5205 | 609.8 | 0.9602 | 161.9 |
| 0.2198 | 398.2 | 0.5605 | 614.7 | | |

Propiophenone (1) + N-methylformamide (2), $T$ = 318.15 K

| | | | | | |
|---|---|---|---|---|---|
| 0.0299 | 71.7 | 0.3705 | 576.9 | 0.7008 | 599.6 |
| 0.0599 | 138.6 | 0.4006 | 596.4 | 0.7308 | 579.3 |
| 0.0899 | 201.1 | 0.4306 | 612.7 | 0.7607 | 551.9 |
| 0.1199 | 258.1 | 0.4606 | 625.9 | 0.7908 | 516.2 |
| 0.1499 | 311.9 | 0.4907 | 635.7 | 0.8208 | 474.4 |
| 0.1799 | 361.5 | 0.5207 | 642.2 | 0.8509 | 419.5 |
| 0.2099 | 407.0 | 0.5508 | 645.2 | 0.8808 | 372.4 |
| 0.2399 | 448.9 | 0.5808 | 644.2 | 0.9105 | 281.1 |
| 0.2698 | 487.2 | 0.6108 | 639.6 | 0.9402 | 205.3 |
| 0.3154 | 533.4 | 0.6409 | 630.8 | 0.9702 | 105.4 |
| 0.3405 | 554.6 | 0.6708 | 618.2 | | |

Propiophenone (1) + N,N-dimethylformamide (2), $T$ = 293.15 K

| | | | | | |
|---|---|---|---|---|---|
| 0.0300 | 14.3 | 0.3598 | 107.2 | 0.6706 | 93.0 |
| 0.0600 | 28.0 | 0.3704 | 107.6 | 0.7006 | 86.8 |
| 0.0900 | 40.4 | 0.4004 | 109.6 | 0.7306 | 81.7 |
| 0.1199 | 51.7 | 0.4305 | 111.1 | 0.7606 | 74.5 |
| 0.1499 | 62.0 | 0.4606 | 111.7 | 0.7905 | 65.9 |
| 0.1799 | 71.2 | 0.4906 | 111.3 | 0.8204 | 60.7 |
| 0.2099 | 79.5 | 0.5206 | 109.5 | 0.8505 | 52.6 |
| 0.2399 | 86.9 | 0.5507 | 108.3 | 0.8802 | 43.4 |
| 0.2699 | 93.4 | 0.5807 | 105.4 | 0.9102 | 32.8 |
| 0.2999 | 99.0 | 0.6106 | 102.1 | 0.9402 | 23.2 |
| 0.3299 | 103.6 | 0.6407 | 97.7 | 0.9701 | 10.9 |

Propiophenone (1) + N,N-dimethylformamide (2), $T$ = 298.15 K

| | | | | | |
|---|---|---|---|---|---|
| 0.0300 | 15.7 | 0.3448 | 115.9 | 0.6707 | 101.4 |
| 0.0600 | 30.5 | 0.3754 | 116.4 | 0.7007 | 95.7 |
| 0.0899 | 44.0 | 0.4005 | 118.3 | 0.7307 | 89.1 |
| 0.1199 | 56.4 | 0.4305 | 119.7 | 0.7606 | 81.9 |
| 0.1498 | 67.8 | 0.4606 | 120.3 | 0.7906 | 74.0 |
| 0.1798 | 78.0 | 0.4906 | 120.1 | 0.8206 | 65.3 |
| 0.2098 | 87.2 | 0.5206 | 118.9 | 0.8505 | 56.0 |
| 0.2398 | 95.3 | 0.5507 | 117.0 | 0.8804 | 45.9 |
| 0.2698 | 102.4 | 0.5807 | 114.3 | 0.9104 | 35.3 |
| 0.2998 | 108.5 | 0.6107 | 110.8 | 0.9401 | 24.1 |
| 0.3298 | 113.7 | 0.6408 | 106.5 | 0.9701 | 12.2 |
| 0.3448 | 115.9 | | | | |

Propiophenone (1) + N,N-dimethylformamide (2), $T$ = 308.15 K

| | | | | | |
|---|---|---|---|---|---|
| 0.0222 | 13.0 | 0.3201 | 121.3 | 0.5996 | 126.4 |
| 0.0449 | 25.6 | 0.3478 | 125.6 | 0.6325 | 121.6 |
| 0.0679 | 37.7 | 0.3628 | 127.7 | 0.6661 | 115.5 |
| 0.0912 | 49.1 | 0.3743 | 130.0 | 0.7003 | 108.4 |
| 0.1150 | 59.9 | 0.3857 | 131.1 | 0.7352 | 99.9 |
| 0.1391 | 70.0 | 0.4146 | 133.3 | 0.7708 | 90.0 |
| 0.1637 | 79.4 | 0.4440 | 134.6 | 0.8069 | 78.8 |
| 0.1887 | 88.2 | 0.4740 | 135.0 | 0.8440 | 66.1 |
| 0.2140 | 96.2 | 0.5045 | 134.4 | 0.8819 | 51.8 |
| 0.2399 | 103.6 | 0.5356 | 132.8 | 0.9203 | 35.9 |
| 0.2661 | 110.2 | 0.5673 | 130.1 | 0.9598 | 18.6 |
| 0.2929 | 116.1 | | | | |

Propiophenone (1) + N,N-dimethylacetamide (2), $T$ = 298.15 K

| | | | | | |
|---|---|---|---|---|---|
| 0.0300 | -8.3 | 0.3598 | -97.9 | 0.6706 | -113.9 |
| 0.0599 | -17.9 | 0.3798 | -101.3 | 0.7006 | -109.6 |
| 0.0898 | -27.5 | 0.4054 | -108.7 | 0.7307 | -104.1 |
| 0.1198 | -36.9 | 0.4303 | -112.3 | 0.7607 | -97.5 |
| 0.1498 | -46.0 | 0.4604 | -115.8 | 0.7906 | -89.7 |
| 0.1798 | -54.9 | 0.4904 | -118.4 | 0.8205 | -80.7 |
| 0.2098 | -63.4 | 0.5205 | -120.2 | 0.8504 | -70.5 |
| 0.2398 | -71.4 | 0.5505 | -120.5 | 0.8803 | -59.0 |
| 0.2698 | -79.0 | 0.5806 | -120.7 | 0.9102 | -46.1 |
| 0.2998 | -85.9 | 0.6106 | -119.5 | 0.9402 | -31.5 |
| 0.3298 | -92.3 | 0.6407 | -117.3 | 0.9701 | -14.9 |

Propiophenone (1) + N,N-dimethylacetamide (2), $T$ = 308.15 K

| | | | | | |
|---|---|---|---|---|---|
| 0.0300 | -9.9 | 0.3898 | -101.9 | 0.6822 | -108.8 |
| 0.0599 | -19.4 | 0.3997 | -103.2 | 0.7116 | -104.3 |
| 0.0898 | -28.9 | 0.4150 | -108.1 | 0.7410 | -98.6 |
| 0.1198 | -38.2 | 0.4433 | -111.5 | 0.7701 | -91.5 |
| 0.1498 | -47.3 | 0.4735 | -114.2 | 0.7991 | -83.8 |
| 0.1798 | -56.1 | 0.5036 | -116.2 | 0.8282 | -74.5 |
| 0.2098 | -64.4 | 0.5336 | -117.2 | 0.8571 | -64.5 |
| 0.2398 | -72.2 | 0.5635 | -117.4 | 0.8859 | -53.6 |
| 0.2697 | -79.4 | 0.5933 | -116.6 | 0.9147 | -41.7 |
| 0.2997 | -86.0 | 0.6231 | -114.9 | 0.9432 | -28.8 |
| 0.3297 | -92.0 | 0.6526 | -112.3 | 0.9718 | -14.2 |
| 0.3597 | -97.3 | | | | |

Propiophenone (1) + N-methyl-2-pyrrolidone (2), $T$ = 298.15 K

| | | | | | |
|---|---|---|---|---|---|
| 0.0299 | -51.1 | 0.3598 | -418.8 | 0.7005 | -372.7 |
| 0.0599 | -102.9 | 0.3898 | -430.7 | 0.7305 | -347.7 |

| $x_1$ | $H^E$ | $x_1$ | $H^E$ | $x_1$ | $H^E$ |
|---|---|---|---|---|---|
| 0.0898 | -150.8 | 0.4403 | -457.2 | 0.7605 | -319.7 |
| 0.1198 | -195.2 | 0.4603 | -459.6 | 0.7904 | -288.6 |
| 0.1498 | -235.8 | 0.4903 | -460.5 | 0.8204 | -254.6 |
| 0.1798 | -272.8 | 0.5204 | -457.9 | 0.8504 | -217.7 |
| 0.2098 | -306.1 | 0.5504 | -451.9 | 0.8804 | -178.0 |
| 0.2398 | -335.8 | 0.5804 | -442.5 | 0.9103 | -135.4 |
| 0.2698 | -361.9 | 0.6104 | -429.9 | 0.9402 | -89.9 |
| 0.2998 | -384.5 | 0.6404 | -414.0 | 0.9702 | -41.0 |
| 0.3298 | -403.4 | 0.6704 | -394.9 | | |

Propiophenone (1) + N-methyl-2-pyrrolidone (2), $T$ = 308.15 K

| $x_1$ | $H^E$ | $x_1$ | $H^E$ | $x_1$ | $H^E$ |
|---|---|---|---|---|---|
| 0.0300 | -58.2 | 0.3898 | -426.1 | 0.6705 | -382.6 |
| 0.0599 | -111.5 | 0.4168 | -432.9 | 0.7005 | -361.7 |
| 0.0899 | -161.0 | 0.4254 | -438.5 | 0.7305 | -337.8 |
| 0.1199 | -205.4 | 0.4404 | -441.3 | 0.7604 | -311.4 |
| 0.1499 | -246.0 | 0.4604 | -443.7 | 0.7904 | -282.0 |
| 0.1799 | -281.8 | 0.4904 | -444.5 | 0.8204 | -249.7 |
| 0.2099 | -313.0 | 0.5205 | -442.1 | 0.8503 | -214.8 |
| 0.2398 | -340.4 | 0.5505 | -436.4 | 0.8804 | -177.2 |
| 0.2698 | -364.1 | 0.5805 | -427.6 | 0.9103 | -137.0 |
| 0.2998 | -384.5 | 0.6106 | -415.6 | 0.9402 | -94.2 |
| 0.3298 | -401.8 | 0.6405 | -400.7 | 0.9700 | -47.1 |
| 0.3598 | -415.6 | | | | |

[a]Standard uncertainties $u$ are: $u(x_1)$ = 1·10$^{-4}$, $u(T)$ = 0.03 K, $u(H^E)$ = 0.005$H^E$ and the combined expanded uncertainty $U_c(H^E)$ = 0.01$H^E$.

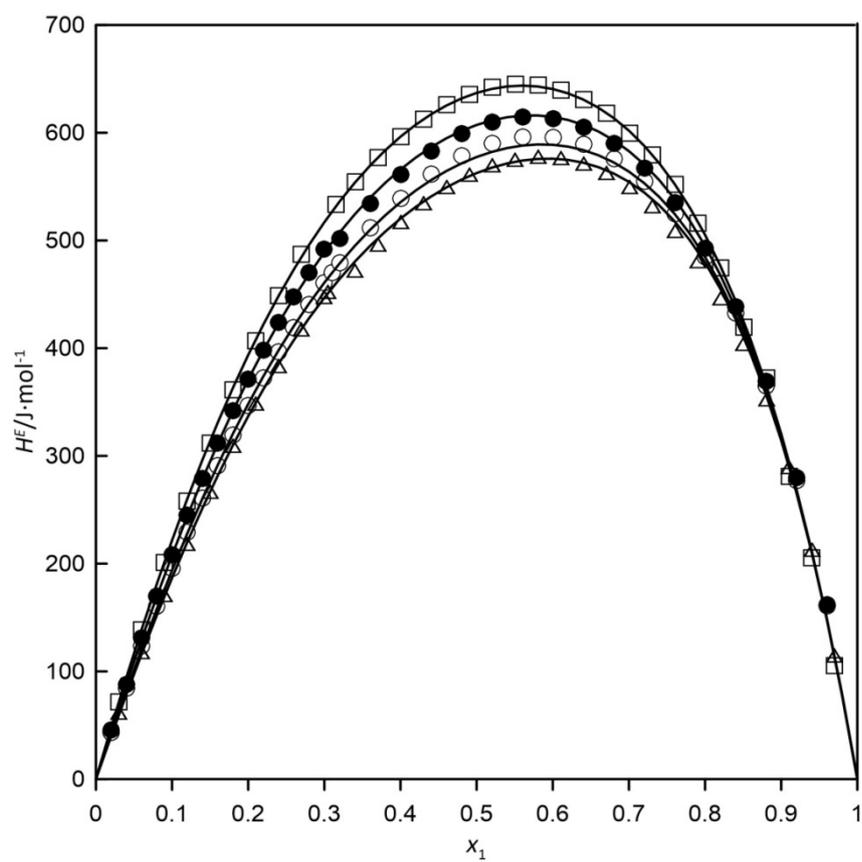

Fig. 6

Excess enthalpies for the propiophenone (1) + N-methylformamide (2) system. Experimental: (Δ) 293.15 K, (ο) 298.15 K, (●) 308.15 K, (□) 318.15 K. Solid lines are calculated by eqn. (8) with the parameters given in Table 6.

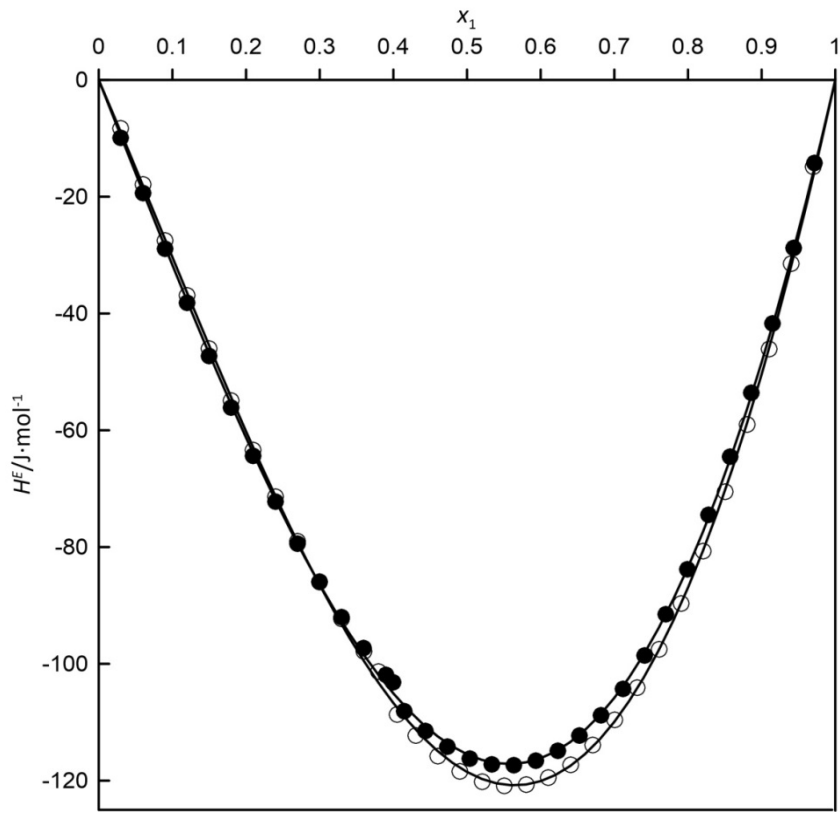

Fig. 7

Excess enthalpies for the propiophenone (1) + N,N-dimethylacetamide (2) system. Experimental: (○) 298.15 K, (●) 308.15 K. Solid lines are calculated by eqn. (8) with the parameters given in Table 6.

## 4. Thermodynamic description

The main idea of combining the solid-liquid equilibrium data with the excess enthalpies was discussed in detail in our previous paper [1]. Here only the most important aspects and conclusions will be shown. The general solubility equation, with some commonly accepted simplifications, has the following form

$$R \ln[x_1 \gamma_1(x_1, T)] = -\Delta H_{m1} \left( \frac{1}{T} - \frac{1}{T_{m1}} \right) \tag{2}$$

where $x_1, \gamma_1, \Delta H_{m1}, T_{m1}$ are mole fraction, activity coefficient, enthalpy and temperature of melting pertaining to the solute, respectively. The practical problem is the representation of activity coefficient which is a simultaneous function of composition and temperature. Any commonly used excess Gibbs energy models, no matter if assuming or neglecting its temperature dependence, determines in fact excess enthalpies if adjusted to the solubility data. The resulting excess enthalpies strongly depend on the model temperature dependence which is optional and can vary within very wide limits. An approach which we have proposed, forces an agreement between solid-liquid equilibrium data and excess enthalpies by using information drawn from experimental excess enthalpies in the adjustment of the activity coefficient function to the solid-liquid equilibrium data.

The activity coefficients are represented in the following form which results directly from the isobaric Gibbs-Duhem equation

$$\ln \gamma_1 = Q + x_2 \left(\frac{dQ}{dx_1}\right) + \frac{x_2 H_m^E}{RT^2}\left(\frac{dT}{dx_1}\right) \tag{4a}$$

$$\ln \gamma_2 = Q - x_1 \left(\frac{dQ}{dx_1}\right) - \frac{x_1 H_m^E}{RT^2}\left(\frac{dT}{dx_1}\right) \tag{4b}$$

where $H_m^E$ is excess enthalpy, $Q = G^E/RT$ and $G^E$ is excess Gibbs energy and all differentiations should be performed along the solubility curve. Combining one from the above equations with eqn. (2) leads to the following solubility equation for the solute 1

$$R \ln x_1 + R\left[Q + x_2 \left(\frac{dQ}{dx_1}\right) + \frac{x_2 H_m^E}{RT^2}\left(\frac{dT}{dx_1}\right)\right] = -\Delta H_{m1}\left(\frac{1}{T} - \frac{1}{T_{m1}}\right) \tag{5}$$

with the analogous form for the component 2 appearing in the solid phase.

The model adjustment concerns the $Q$-function which should be defined as an explicit function of mole fractions only. The concentration is dependent on temperature along the solubility curve and thus the temperature dependence of the $Q$-function is also acknowledged.

## 5. Calculations

The $Q$-function was described by the Redlich-Kister type power expansion, i.e.

$$Q = G_m^E/RT = x_1 x_2 \sum_{i=0}^{n} a_i(x_1 - x_2)^i \tag{6}$$

with the $a_i$ parameters being independent both of $x_1$ and $T$. The final equation of the activity coefficients is following

$$\ln \gamma_1 = x_2^2\left[\sum_{i=0}^{n} a_i(x_1 - x_2)^i + 2x_1 \sum_{i=1}^{n} i a_i(x_1 - x_2)^{i-1}\right] + \frac{x_2 H_m^E}{RT^2}\left(\frac{dT}{dx_1}\right) \tag{7a}$$

$$\ln \gamma_2 = x_1^2\left[\sum_{i=0}^{n} a_i(x_1 - x_2)^i - 2x_1 \sum_{i=1}^{n} i a_i(x_1 - x_2)^{i-1}\right] - \frac{x_1 H_m^E}{RT^2}\left(\frac{dT}{dx_1}\right) \tag{7b}$$

Also the Redlich-Kister equation with the polynomial temperature dependence of parameters was used to correlate experimental excess enthalpies

$$H_m^E = x_1 x_2 \sum_{i=0}^{n} \sum_{j=0}^{m} b_{ij}(T/T_0)^j (x_1 - x_2)^i \tag{8}$$

with the reference temperature $T_0$ equal to 298.15 K.

The number of the expansion terms ($n$+1) and ($m$+1) in eqns. (6) and (8) was determined by successive approximations with the increasing $n$ and $m$ number. If the further increase did not cause a significant decrease in the standard deviation, the procedure was stopped and the previous number accepted as optimal. For three systems – propiophenone + {NMF, or DMA, or NMP} the linear temperature dependence turned out to be sufficient ($m$ = 1) while the system with DMF the quadratic relation was assumed ($m$ = 2) – see Table 6.

In the beginning, the $b_{ij}$ parameters of eqn. (8) were determined by the adjustment to experimental excess enthalpies. Next, the $a_i$ parameters of eqn. (6) were adjusted to the solubility temperatures, using eqns. (2) and (7). The objective function was defined as follows

$$F(a_0, a_1, \dots, a_n) = \sum_{i=1}^{m}\left[T_i^{exp} - T^{calc}\left(x_{1i}^{exp}; a_0, a_1, \dots, a_n\right)\right]^2 \tag{9}$$

where superscripts *exp* and *calc* denote experimental and calculated value of a parameter corresponding to the *i*th experimental point and the summation is performed over all *m* experimental points. The calculated temperature ($T^{cal}$) is the solution of the solubility equation – eqn. (2), with respect to the temperature for the experimental concentration of saturated solution.

The solubility equation interrelating $T$ and $x_1$ and being a combination of eqns. (2) and (7) is in fact a differential equation, because of the derivative $\left(\frac{dT}{dx_1}\right)$ appearing therein. Formally it can be solved iteratively as with known or assumed values of model parameters ($a_i$, $b_{ij}$) this derivative can be calculated numerically. However, to simplify calculations, we used a different approach. The solubility data were separately correlated by a following empirical equation called the Apelblat equation [12]

$$\ln x_i = \frac{A}{T/K} + B \ln(T/K) + C \tag{10}$$

and the derivative $\left(\frac{dT}{dx_1}\right)$ was calculated using the above equation. In the case of insufficient number of data, the simplified form with $B = 0$ was used. The values of the adjusted parameters with the standard deviations of temperature are given in Table 5.

Table 5. Correlation of the solubility data by means of the Apelblat equation - eqn. (10).[a]

| System | $10^{-3} \cdot A$ /K | $B$ | $C$ | $\sigma^{b}$ /K |
|---|---|---|---|---|
| Propiophenone + | 21.052 | 98.749 | -632.54 | 0.5 |
| N-methylformamide | -1.5398 | 0 | 5.6902 | 0.3 |
| Propiophenone + | -2.2106 | 1.4904 | -86.317 | 0.7 |
| N,N-dimethylformamide | | | | |
| Propiophenone + | -0.74381 | 5.4353 | -28.312 | 0.4 |
| N,N-dimethylacetamide | 5.0174 | 23.046 | -147.32 | 0.4 |
| Propiophenone + | 1.0149 | 11.060 | -66.270 | 0.6 |
| N-methyl-2-pyrrolidone | 126.87 | 523.26 | -3396.6 | 0.3 |

[a] The name of the component in the leftmost column indicates the solid phase for the solubility curve with parameters written on the right. [b] Standard deviation $\sigma = \left[\frac{1}{m}\sum_{i=1}^{m}\left(T_i^{exp} - T^{calc}\left(x_{1i}^{exp}\right)\right)^2\right]^{1/2}$

The values of adjusted parameters to the excess enthalpies and solid-liquid equilibria data are grouped in Table 6. Also standard deviations of the fit are shown therein.

Table 6. Correlation of the excess enthalpy and solid-liquid equilibrium data by the Redlich-Kister equation (6, 8).

| System Propiophenone + | $i$ | $10^{-3}b_{i0}$ /J·mol$^{-1}$ | $10^{-3}b_{i1}$ /J·mol$^{-1}$ | $10^{-3}b_{i2}$ /J·mol$^{-1}$ | $\sigma^a$ /J·mol$^{-1}$ | $a_i$ | $\sigma^b$ /K |
|---|---|---|---|---|---|---|---|
| N-methylformamide | 0 | -1239 | 3549 | | 3.7 | 1.650 | 0.44 |
| | 1 | 1879 | -1343 | | | 0.1366 | |
| | 2 | 2122 | -1308 | | | 0.3150 | |
| | 3 | 2114 | -1629 | | | | |
| N,N-dimethylformamide | 0 | -15.51 | 29.86 | -13.88 | 0.65 | 0.2903 | 0.65 |
| | 1 | 17.46 | -34.87 | 17.34 | | | |
| N,N-dimethylacetamide | 0 | -835.6 | 361.6 | | 0.66 | -0.5073 | 0.60 |
| | 1 | -848.1 | 708.8 | | | 0.4703 | |
| | 2 | 426.5 | -386.8 | | | -0.2796 | |
| | | | | | | 1.008 | |
| | | | | | | -0.8598 | |
| N-methyl-2-pyrrolidone | 0 | -3636 | 1805 | | 2.4 | -0.1463 | 0.51 |
| | 1 | -1901 | 1986 | | | 1.301·10$^{-3}$ | |
| | 2 | 5959 | -5835 | | | 0.3964 | |

[a,b] Standard deviations: (a) $\sigma = \left[\frac{1}{m}\sum_{i=1}^{m}\left(H_i^{E\ exp} - H^{E\ calc}\left(T_i, x_{1i}^{exp}\right)\right)^2\right]^{1/2}$; (b) $\sigma = \left[\frac{1}{m}\sum_{i=1}^{m}\left(T_i^{exp} - T^{calc}\left(x_{1i}^{exp}\right)\right)^2\right]^{1/2}$

We checked the reliability of the prediction of measured properties using the modified UNIFAC model [13]. However, there is a problem with the availability of the necessary group parameters. If the idea of the group definition as applied in the UNIFAC approach is to be used, the carbonyl group adjacent to the aromatic ring in the propiophenone molecule should be considered as a separate group with unique group parameters. Unfortunately, such parameters have not yet been determined due to the lack of sufficient experimental data, as can be guessed. The only practical possibility seems to be to ignore the differences between the carbonyl group adjacent to the aromatic part and the aliphatic part. Finally we adopted the parameters of the CH2CO (aliphatic) group as a possible estimate for the carbonyl group adjacent to the aromatic ring. The division into the groups is shown in Table 7. While this approach can be considered as a crude simplification, we believe that it allows for a qualitative estimation of the differences between properties of different systems with a common component (i.e. propiophenone). The modified UNIFAC model [13] with the group parameters developed by the NIST group [14] was used.

Table 7. The NIST-Modified UNIFAC group assignment [14].

| UNIFAC group | Sub-group number [14] | Number of groups in a molecule of | | | | |
|---|---|---|---|---|---|---|
| | | Propiophenone | NMF | DMF | DMA | NMP |
| CH3 | 1 | 1 | | | 1 | |
| ACH | 9 | 5 | | | | |
| AC | 10 | 1 | | | | |
| CH2CO | 18 | 1 | | | | |
| DMF | 72 | | | 1 | | |
| c-CH2 | 78 | | | | | 3 |
| c-CON-CH3 | 86 | | | | | 1 |
| HCONHCH3 | 93 | | 1 | | | |
| CON(CH3)2 | 101 | | | | 1 | |

## 6. Results and discussion

All the systems exhibit solid-liquid equilibria in the form of a simple eutectic with highly unsymmetrical two liquidus solubility curves (Table 3, Fig. 2-5). This lack of symmetry is a direct consequence of significant differences between temperatures of melting of pure compounds. That of propiophenone (291.4 K) is considerably higher than of other components.

The number of the Redlich-Kister parameters necessary to reproduce experimental solubility temperatures with the standard deviation close to experimental uncertainty varies from one to five. The values of parameters and standard deviations are given in Table 6. The system with NMF exhibits high positive deviations from ideality what manifests itself in solubilities lower than ideal ones. In the series of remaining components, these deviations decrease, still being positive for DMF, almost null for DMA and finally significantly negative for the propiophenone + NMP mixture.

The quality of prediction of the solid-liquid equilibrium data by the modified UNIFAC model is ambiguous. For the systems with NMF and NMP, it brings a significant improvement over ideal solubility although its quality can be regarded as merely qualitative (Fig. 2, 5). For the propiophenone + DMF system, negative deviations from ideality are predicted and positive for the system with DMA while in both cases just opposite property is observed. Both systems exhibit relatively low deviations from ideality. This is a state where the low excess Gibbs energy is due to summing up large and opposite contributions. In these two cases, neglecting the solute activity coefficient gives better results than the modified UNIFAC model.

A similar pattern as for solid-liquid equilibria is observed for the excess enthalpies. The observed heat effects decrease in the order NMF > DMF > DMA > NMP, for the first two being positive and negative for the remaining ones. This observation suggests that deviations from ideality have mainly enthalpic character.

For the propiophenone + N-methylformamide system, the excess enthalpies were measured at four temperatures – (293.15, 298.15, 308.15, 318.15) K. For the remaining systems, the measurements were performed at 298.15 K and 308.15 K, only. The temperature dependence of excess enthalpy is qualitatively similar for all the systems studied. The excess enthalpy increases with the increasing of the temperature, that is, the excess heat capacity is positive. Measurements performed at four temperatures for the propiophenone + NMF system, made possible an estimation of the excess heat capacity through the derivation of eqn. (8) with respect to the temperature. The resulting expression for the excess heat capacity has the following form

$$c_{pm}^E = x_1 x_2 \sum_{i=0}^{n}(b_{i1}/T_0)(x_1 - x_2)^i \qquad (11)$$

and the calculated dependence is shown in Figure 8. The positive and relatively high maximum value is typical for the systems with high positive deviations from ideality and strong intermolecular interactions – see for example [15].

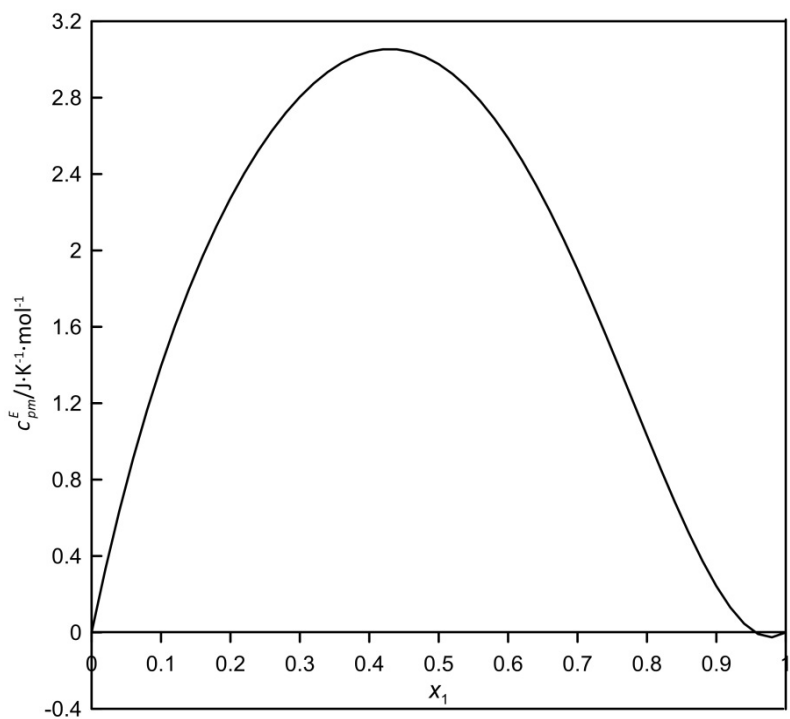

Fig. 8

Excess heat capacity for the propiophenone + N-methylformamide system calculated on the base of the excess enthalpy data at (293.15, 298.15, 308.15, 318.15) K.

The Redlich-Kister correlation equation, i.e. eqn. (8) requires eight (for the system with NMF) or six adjustable parameters to represent the excess enthalpy data having experimental combined uncertainty equal approximately 1%. It is a quite low number if compared with the described experimental data points– 133 for the system with NMF, 100 for the system with DMF and about 65 for the remaining ones (Table 6).

As no excess enthalpy data are available for mixtures of propiophenone with amides, the only possible comparison is with data of similar systems containing acetophenone. The differences between excess enthalpies of the systems containing propiophenone and acetophenone are more significant than for the solid-liquid equilibrium data. Table 8 shows the extremum excess enthalpy values for both series of the data. The similar trend, i.e. decreasing of the excess enthalpy in the order NMF > DMF > DMA > NMP is observed for both of them. The systems with propiophenone have excess enthalpies shifted towards higher, that is more endothermic, values. It can be explained by enlargement of the aliphatic chain in the propiophenone molecule. Indeed, the enthalpies of vaporization at the temperature of about 400 K are almost identical for both compounds and equal about 52 kJ·mol$^{-1}$ [16] what leads to the significantly lower value for propiophenone if related to unit volume or molar mass.

Table 8. The extremum values of experimental excess enthalpies for the systems containing acetophenone and propiophenone at 308.15 K.

|  | Acetophenone[a] (1) + | | Propiophenone (1) + | |
| --- | --- | --- | --- | --- |
|  | $x_1$ | $H^E$/J·mol$^{-1}$ | $x_1$ | $H^E$/J·mol$^{-1}$ |
| + N-methylformamide | 0.5807 | 479.8 | 0.6005 | 613.1 |
| + N,N-dimethylformamide | 0.4904 | -127.9 | 0.4606 | 120.3 |
| + N,N-dimethylacetamide | 0.5205 | -375.9 | 0.5635 | -117.4 |
| + N-methyl-2-pyrrolidone | 0.4902 | -722.0 | 0.4904 | -444.5 |

[a]Data taken from ref. [1].

The excess enthalpy prediction by the modified UNIFAC at 298.15 K is illustrated in Fig. 9. For the systems with NMF, DMA and NMP, the results are surprisingly good and unacceptable only for the system with DMF for which negative excess enthalpy is predicted while actually a low endothermic effect is observed. It is worth noting that for the similar systems with acetophenone, the overall results of the modified UNIFAC application were considerably worse [1].

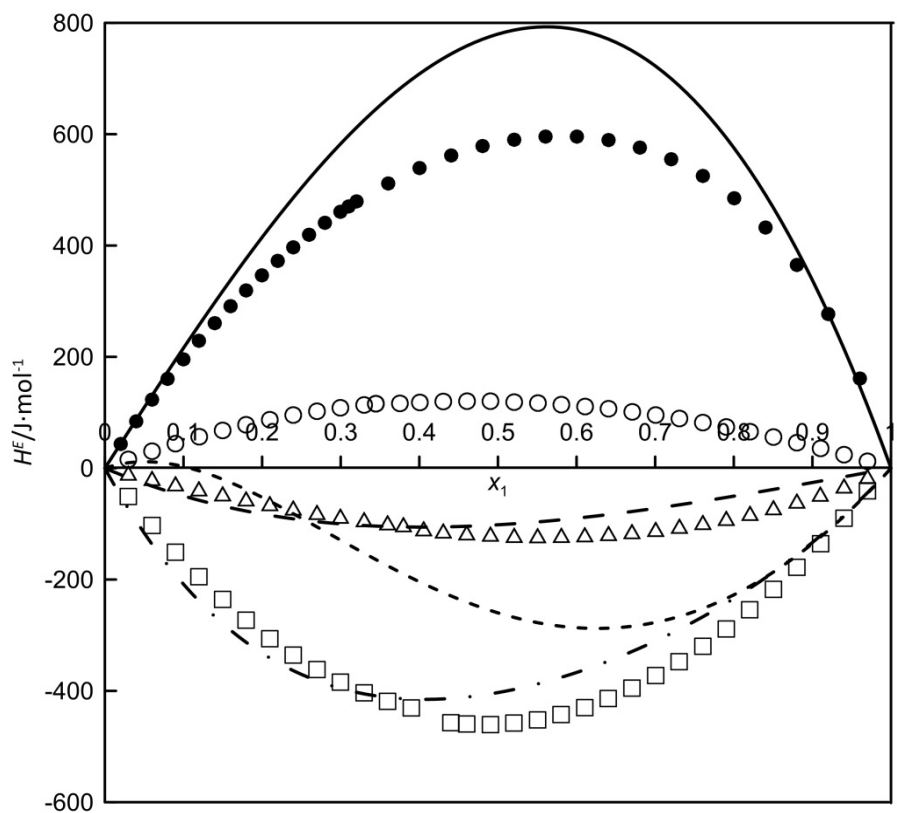

Fig. 9

Experimental excess enthalpies (symbols) and predicted by the modified UNIFAC model (lines) at 298.15 K for the system propiophenone + (●, solid) N-methylformamide, (○, small dash) N,N-dimethylformamide, (△, long dash) N,N-dimethylacetamide, (□, dash + dot) N-methyl-2-pyrrolidone.

The properties of mixtures with N-methylformamide differ significantly from other systems. They exhibit high positive deviations from ideality what causes a significant decrease of solubility. The excess enthalpies are also positive and high. This behavior has been usually explained by selfassociation of NMF molecules [17] [18]. It is confirmed by the excess enthalpy as a function of composition which is more unsymmetrical than for other mixtures.

Similarly as for the systems with acetophenone, negative deviations from ideality for the mixtures with DMA and particularly NMP, and simultaneously negative excess enthalpies are observed. The extent of these effects is less pronounced than in mixtures with acetophenone, but still is not quite understood. Negative deviations from ideality having mainly enthalpic character are usually explained by strong specific attractive interactions between unlike molecules. It is not clear, however, what particular interactions between unlike molecules can predominate over those between like ones. It should be noted, that a similar trend as for propiophenone + {DMF, +DMA, + NMP} is observed for the excess enthalpies for the binary systems of benzene + {DMF, +DMA, + NMP} [19] [20]. This suggests that there are strong interactions between the aromatic ring and the amide group and that they are mainly responsible for the observed negative deviations from ideality.

## 8. Conclusions

The data of solid-liquid equilibrium and excess enthalpy for the systems of propiophenone and {N-methylformamide, or N,N-dimethylformamide, or N,N-dimethylacetamide, or N-methyl-2-pyrrolidone) were presented. They were correlated by the equation based on the Redlich-Kister approach with standard deviations consistent with experimental uncertainties. The excess enthalpies were described by the Redlich-Kister equation with its parameters dependent on temperature. In the description of solubilities, the excess enthalpy as a function of concentration and temperature was used to account for the temperature dependence of the solute activity coefficient.  The properties of the propiophenone + NMF system turned out to be outstanding, with high deviations from ideality and positive excess enthalpies, and consequently with relatively low solubility. For the remaining systems, the deviations from ideality changes into negative and with exothermic excess enthalpy, having the highest extent for the propiophenone + NMP mixture. All  measured properties show a systematic shift compared to analogous acetophenone data as a consequence of the enlargement of the aliphatic part of the propiophenone molecule.

An application of the modified UNIFAC in the prediction of the measured data gave fairly good results for some systems, better than expected if crude simplifications assumed in the definition of structural groups building the propiophenone molecule are recalled. The excess enthalpies are predicted with semi-quantitative accuracy with an exception of the propiophenone + DMF system. The solubilities for two systems – propiophenone + {NMF or NMP} are better predicted by the modified UNIFAC model than application of the ideal solubility approach.

### 9. Acknowledgements

The authors gratefully acknowledge the financial support received from the Consejería de Educación y Cultura de Junta de Castilla y León, under Project VA100G19 (Apoyo a GIR. BDNS:425389). A. Cobos is grateful to Ministerio de Educación, Cultura y Deporte for the grant FPU15/05456 and for the complementary grant EST16/00839.

This work was partly financially supported by Warsaw University of Technology.

**Notes**

The authors declare no competing financial interests.